# Brillouin imaging in turbid samples: the removal of multiple scattering contribution

Maurizio Mattarelli[1]*, Giulio Capponi[1], Alessandra Anna Passeri[1], Daniele Fioretto[1,2], Silvia Caponi[3]*

[1] Dipartimento di Fisica e Geologia, Università di Perugia, Via A. Pascoli, I-06100 Perugia, Italy

[2] CEMIN, Centre of Excellence on Nanostructured Innovative Materials, University of Perugia, Via Elce di Sotto 8, 06123 Perugia, Italy.

[3] Istituto Officina dei Materiali, Italian National Research Council (IOM-CNR), Unit of Perugia, c/o Department of Physics and Geology, University of Perugia, Via A. Pascoli, I-06123 Perugia, Italy

*corresponding authors: silvia.caponi@cnr.it, maurizio.mattarelli@unipg.it

Abstract

We provide a new analytic expression and an innovative experimental method to isolate the effect of multiple scattering (MS) in Brillouin investigation of highly turbid media. On the one hand, an analytic model is given to describe the spectrum in case of ill-defined exchanged wave-vector. On the other hand, a new experimental method, named Polarization Gated Brillouin Spectroscopy (PG-BS), is proposed for selecting the MS contribution through light polarization. Both experimental and analytic methods are tested against a benchmark material, milk, demonstrating their capability to extract reliable micro-mechanical parameters even in highly turbid materials till now inaccessible to in-depth Brillouin scattering investigation.

In recent years, the biomedical community has realized the crucial link between mechanical properties and functionality of cells and tissues. Evidences motivated the researchers on the one hand to disclose the molecular origin of mechanical alterations [1], and on the other hand to develop new diagnostic tools based on biomechanical characterization [2]. Different optical methods have already made the transition from the fundamental research to clinical practice with significant impact on medical field. In this framework, Brillouin microscopy (BM) is emerging as a new optical elastography technique [3,4]: exploiting the light-matter interaction, BM provides non-contact mechanical maps of heterogeneous materials at the microscale adding a new contrast parameter in bioimaging [5,6]. Measuring the frequency shift, $\omega_B$, of the light inelastically scattered by spontaneous acoustic phonons, BM provides the longitudinal elastic modulus through the relation $M=\rho\omega_B^2/q^2$ being $\rho$ the density and **q** the exchanged wavevector in the scattering event [3,4]. Promising applications in biomedical arena [3,4,7–9], forced the development of innovative technological solutions to increase acquisition speed and spatial resolution. New Brillouin devices became compatible with the in-vivo diagnosis [7] and reached the subcellular spatial resolution [10–13]. To further enlarge the technique application field toward endoscopy [14] or to deep 3D tissues imaging, BM has to face a challenging stumbling block: the characterization of objects immersed inside turbid media. To achieve this goal, significant limitations on both experimental and theoretical side must be overcome. In fact, when the light crosses a turbid medium characterized by local heterogeneities in the refractive index, it undergoes multiple elastic scattering (MS) losing its initial characteristics in terms of propagation directionality, coherence and polarization [15,16]. As occurring in many optical characterizations, this process has strong impact also in Brillouin spectroscopy. From a fundamental point of view, **q** is the key parameter providing the link between the characteristics of the scattered light and the sample properties. In highly turbid media and in general in optically heterogeneous samples, the module and direction of **q** are no longer univocally defined by the external optical settings (Fig. 1a and Fig. 1b). As a consequence, the Brillouin spectral shape is modified and the **q** uncertainty may transfer into errors in the determination of the sample viscoelastic parameters. Only a specific data treatment and interpretation can extract the actual mechanical properties of the sample without falling into evaluation mistakes.

From the experimental side, detecting the tiny Brillouin signal under a huge elastic spectral tail is a long-standing obstacle for analysis of biological tissues, powders, colloidal solutions or aggregates. The presence of intense elastic scattering in turbid media limits the Brillouin applications mainly to transparent materials. The key instrumental factor to get access to tiny inelastic signals against strong elastic line is the spectral contrast of the spectrometer i.e. the ratio between the maximum and the minimum of the elastic transmission. Here we used the recently developed High Contrast Tandem Fabry-Perot interferometer, HC-TFP, which combines unprecedented contrast -greater than $10^{15}$- to high spectral resolution (~100 MHz) [17], achieving an excellent data quality with negligible elastic background even in turbid samples. The light of a 532 nm single-mode solid-state laser beam (SpectraPhysics Excelsior) is focalized and collected by a Mitutoyo M-Plan Apo 20× with working distance of 20 mm and numerical aperture of 0.42. The presence of a polarizer on the optical path let us to select the polarization of the scattered light. We used as a benchmark sample, a commercial partially skimmed cow milk (1,6% fat content). Milk is a complex biological fluid composed by an emulsion of fat globules (> 1 µm) and a suspension of protein micelles (50-500 nm). Their presence induces MS, which is at the origin of the white color, and produces important consequences in any optical based analysis [18]. In non-absorbing media, the parameter which provides an estimation of the MS amount is the optical density, $OD = l \cdot \mu_e$, where, $\mu_e$ is the extinction coefficient of the material and $l$ the optical path [19]. The sample under study was characterized by spectrophotometry giving $\mu_e$= 20 mm$^{-1}$ at 532 nm corresponding to a scattering length of 50 µm. The back-scattering configuration here exploited, allows us to increase OD by focusing light deeper and deeper inside the sample. Milk, loaded into a quartz cuvette of 10 mm, is mounted on a three-axes piezo-translation stage. A linear scan of the focal position was performed, starting from the sample surface and going in depth into the sample until reaching 650 µm. At this depth, the OD is more than 10 and the relative amount of photons elastically diffused at each angle is expected to be uniform, losing any preferential direction with respect to the initial one [19,20]. The sequence of the Brillouin spectra acquired at different depths, $l$, is reported in Fig. 1c. They show a continuous growth of the low frequency side of the Brillouin peak accompanied by a progressive shift towards lower frequency. The most common data analyzes (DHO fit [21], Lorentzian fit [4], or momentum analysis [22]) read these spectral modifications as a variation in the Brillouin frequency position and width. However, milk is a homogenous material, which, independently from the focusing depth, is characterized by a well-defined elastic modulus, therefore the calculated variations are only spurious effects due to the presence of MS.

As shown in Fig.1b, MS is responsible for inelastic photon-phonon scattering events occurring at all possible values of q. In epidetection, this leads to the actual decrease of the effective exchanged wave-vector, which assumes its maximum value in back-scattering configuration. We modelled the MS effect on the spectra assuming that the total Brillouin intensity is the sum of two contributions, namely $I_{BS}$ and $I_{MS}$. $I_{BS}$ is the Brillouin intensity due to photons whose trajectory is unaffected by the optical heterogeneities - the so called ballistic photons. $I_{MS}$ is the Brillouin intensity due to the multiple scattered - or diffusive - photons.

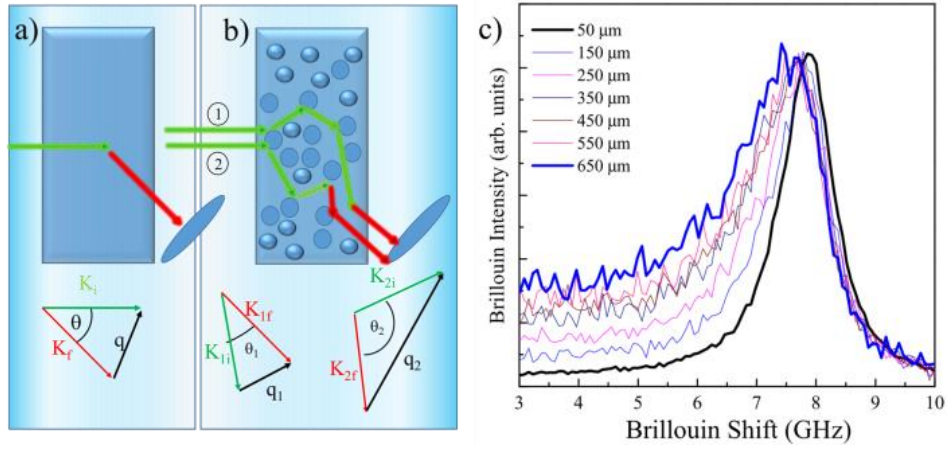

**Figure 1:** Loss of the definition of the exchanged wavevector, $q = \frac{2\pi n}{\lambda}\sin\frac{\vartheta}{2}$, being $n$ the refractive index, $\lambda$ the wavelength of the incident beam and $\vartheta$ the scattering angle, as the analysis moves from transparent (a) to turbid (b) material. c) Evolution of Brillouin peak investigating milk at different penetration depths.

For a quantitative evaluation, we started from the theoretical derived expression of the Brillouin line shape of transparent viscoelastic materials, which at a single q is the DHO function:

$$I(\omega) = \frac{I_0}{\pi}\frac{\omega_B^2 \Gamma}{(\omega^2 - \omega_B^2)^2 + \omega^2 \Gamma^2} \qquad (1)$$

It includes 3 fitting parameters: the frequency position, $\omega_B$, the width, $\Gamma$, and the integrated intensity of the Brillouin peak, $I_0$. Far away from relaxation processes, $\omega_B$ is linearly linked to $q$ through the sound velocity of the acoustic waves, $v$, i.e. $\omega_B = vq$, while $\Gamma$, has a $q^2$ dependence through the kinematic viscosity, $D$, i.e. $\Gamma = Dq^2$. The $q$ variation induced by MS is consequently transferred on modifications of the shape of Brillouin spectra. In presence of a distribution of scattering wave-vectors and considering the $q$ dependences of $\omega_B$ and $\Gamma$, we can write the scattered intensity as:

$$I(\omega) = I_{BS} + I_{MS} \qquad (2)$$

$$I_{BS} = A_{BS}\int_{Q_{obj}} \frac{(vq)^2 Dq^2}{(\omega^2 - (vq)^2) + \omega^2 (Dq^2)^2} R(\mathbf{q})\, d\mathbf{q}$$

$$I_{MS} = A_{MS}\int_0^{q_{BS}} \frac{(vq)^2 Dq^2}{(\omega^2 - (vq)^2) + \omega^2 (Dq^2)^2} R_{MS}(q)\, dq$$

where $v$, $D$, $A_{BS}$ and $A_{MS}$ are fitting parameters. In Eq. 2, the first integral, already introduced in ref. [13], takes into account the $\mathbf{q}$ spread related to the numerical aperture of the microscope objective in the chosen scattering geometry. The second term in Eq. 2 considers the $q$ spread induced by the MS process. We have modeled the weight function, $R_{MS}(q)$, considering that MS induces an isotropic distribution of the scattered radiation over the whole solid angle. In this condition, $R_{MS}(q) \propto q$ results from simple trigonometric relations and Eq.2 has a closed form (see supplementary materials), well suited to be used in the fitting procedure.

The collected spectra were analyzed using Eq. 2 and the results are reported in Fig. 2. Excellent agreement was found between the fitting function and the data at any investigated depth (Fig. 2a) regardless of the MS amount. In fact, all the spectra are well described by the sum of the two specified components, with increasing weight of the MS component moving deeper inside the sample. This is an important advance with respect to the few attempts till now proposed [23,24]. In particular, the data analysis proposed in ref. [24], which is based on an arbitrary choice of the fitting interval, can be applied only in low turbidity conditions and in back scattering geometry. Moreover, the q distribution function proposed to analyze the spectra in ref. [23], which is again valid for low turbid samples, has been arbitrarily selected among several functions of similar shape that met qualitative physically sound assumptions.

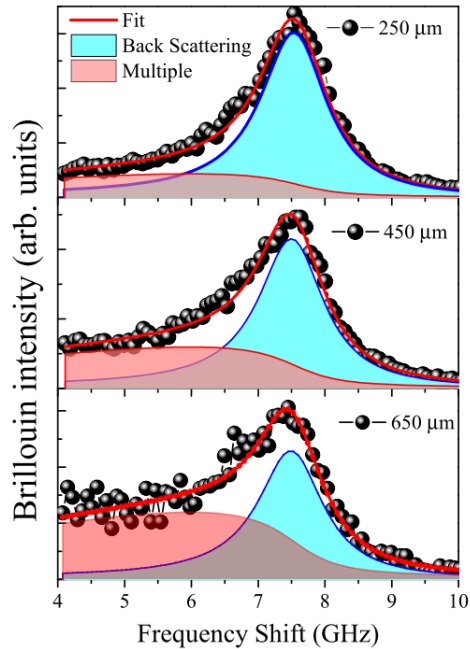

**Figure 2**. a) Brillouin peak (dots) and fitting curve (red line) using eq (2) for three selected focusing depth. The two contributions back scattered (blue line) and multiple (pink) contributions are reported.

The generalization of Eq. 2 to the case of elastically heterogeneous and/or anisotropic samples, typical condition of biological materials, would lead to a much more complicated expression, dependent on the local sample morphology and on the elastic properties of all the different regions, crossed by the photons. To overcome this problem, we propose a complementary approach where the MS component can be directly measured by polarization analysis.

In transmission imaging, ballistic and multiple scattering light components are experimentally separated on the basis of the arrival time on the detector. In fact, the diffusive contribution takes longer time to exit the sample with respect to the ballistic one [15,25]. In the steady state configuration of Brillouin scattering this solution is not applicable. However, as already proved in transmission imaging, also light polarization can be an effective parameter to isolate photons originated from different processes [26]. In fact, the spatial refractive index fluctuations, related to local modifications in composition or density, induce a loss in the memory of the direction of the incident/scattered beam, as well as in its polarization [15]. With this in mind, we developed and tested an innovative experimental strategy, called Polarization Gated Brillouin Spectroscopy (PG-BS), able to correct the Brillouin spectra affected by MS.

In isotropic bulk materials, the Brillouin scattering process produced by longitudinal acoustic waves induces a frequency shift of light while maintaining its polarization. The depolarized Brillouin component, which is originated by shear waves in solid materials and by DID and the re-orientational modes of optically anisotropic molecules in fluids, has a null or negligible intensity in backscattering configuration [27]. On the contrary, the Brillouin spectra acquired in turbid media should present a mixed polarization character, as only the Brillouin signal due to the ballistic photons maintains the polarization of the incoming laser beam, while the one due to the diffuse photon presents also a depolarized component.

To verify this hypothesis and to quantify this effect, we acquired at different focal depths both unpolarized (without selecting any polarization in the scattered light) and depolarized spectra. As a matter of fact, entering inside the samples, the depolarized component is practically constant in intensity and its spectral shape is invariant with the penetration depth as shown in Fig. 3 b). Indeed, apart from the negligible effect of re-orientational modes, this spectrum is the experimental determination of the Brillouin scattering contribution due to the diffusive photons, which have lost the memory of their initial polarization. In Fig. 3a, the comparison between unpolarized (open circles) and depolarized (green line) spectra are reported. They are normalized in the low frequency region, where the spectral intensity can be assigned, as a first approximation, to the diffusive contribution. The subtraction of the depolarized spectrum form the unpolarized one gives us the Brillouin spectra free from the MS contribution. The so obtained data are reported for selected depths in Fig. 3a and for all the investigated focal penetration depths in Fig. 3c. They show the same spectral shape: the spectrum taken at the surface, which is only marginally affected by MS, is scarcely distinguishable from the spectrum obtained at high depth in which the ballistic and diffusive contributions are actually comparable.

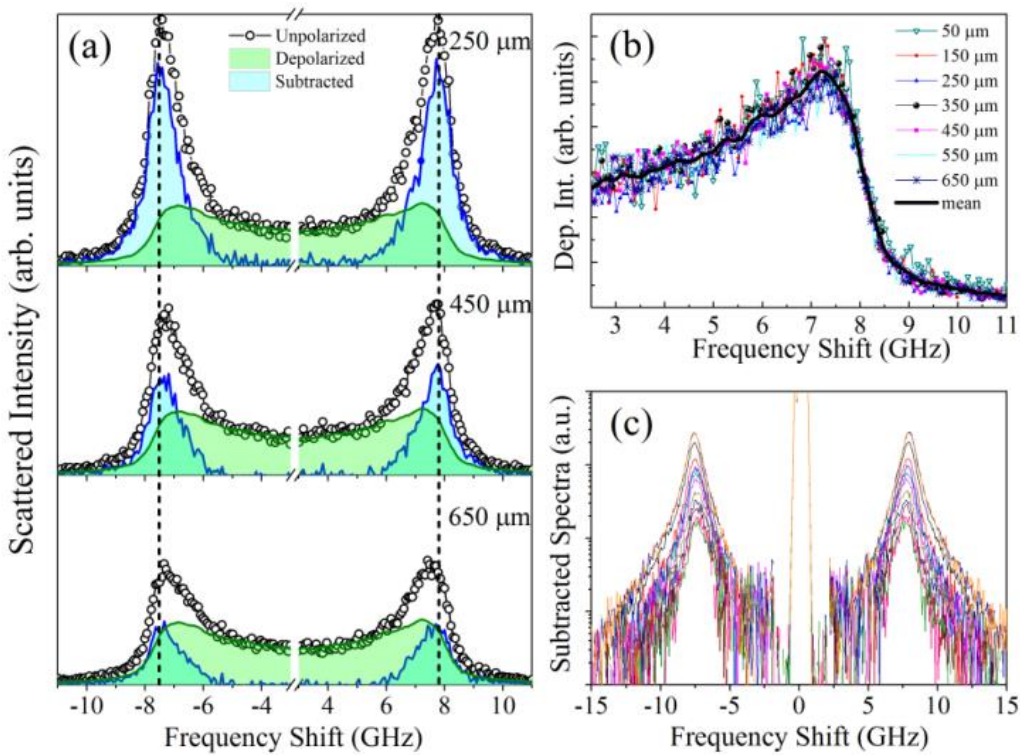

**Figure 3: a)** Unpolarized (open circles) and depolarized (green line) spectra acquired on milk at the declared focusing position inside the sample. Their difference is reported as blue line. b) Normalized VH spectra acquired as a function of the penetration depth. c) Brillouin spectra depurated from the diffusive contribution. They are obtained by the subtraction of depolarized from unpolarized spectra.

To quantitatively compare the results reachable by the two newly proposed approaches and to quantify their reliability in comparison with the most common fitting models, we report in Fig. 4 a) and b) the relative variation of the viscoelastic parameters as a function of the focal penetration depth, $l$. While most common fitting models (DHO fit [21], Lorentzian fit [4], or momentum analysis [22]) find a variation up to 15% in the Brillouin frequency position and of about 300% in the width, there is a drastic reduction in the depth dependence when the here proposed methods are used. This result highlights the ability of both methods to correct the spurious MS effect in the spectra. In independent way, they deliver values of $v$ and $D$ in excellent agreement validating the present model for the Brillouin scattering processes occurring in turbid media. The small residual $l$ dependence, scarcely affecting the results, evidences the presence of a further minor contribution not completely taken into account by our analyses. It can be assigned to the Brillouin scattering events occurring to the photons only slightly deviated from their original trajectory, the so-called snake photons [15,25]. They partially maintain their directionality and polarization. In fact, this contribution is safely neglected also in optical imaging [15,26]. Here, estimating the Brillouin contributions due to the diffusive and ballistic photons, we give the foremost correction, which is enough to provide unprecedented results in highly turbid media analysis.

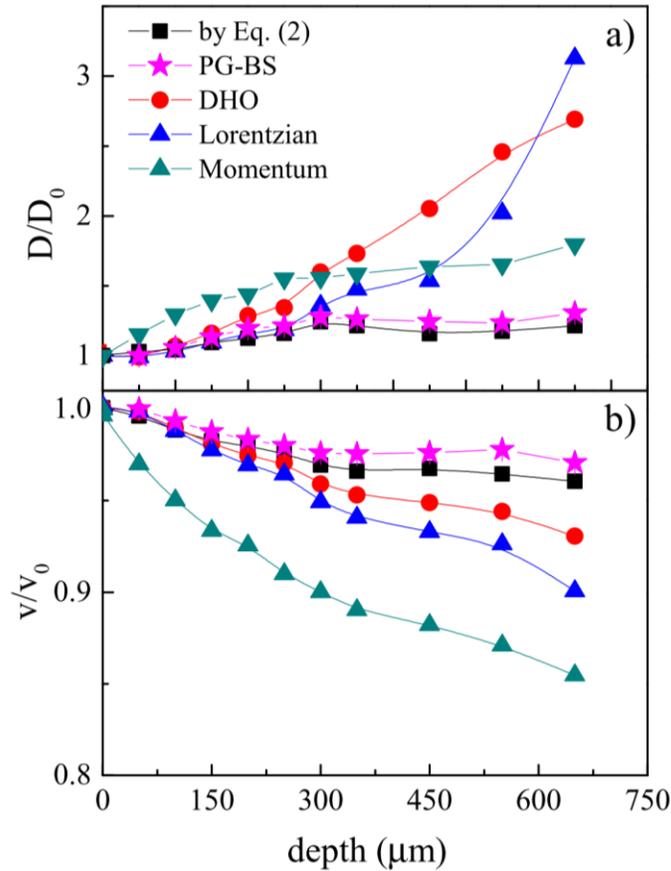

**Figure 4**: Relative variation of sound velocity a) and Brillouin width b) with respect to the value at the surface. The values are obtained using analytic fitting procedure of Eq (2) (squares), the experimental procedure PG-BS method (stars), DHO fit (circles), Lorentzian fit (up-triangles) and momentum analysis (down triangles).

In conclusion in this letter, we have: i) presented a detailed experimental characterization of Brillouin spectra in presence of MS; ii) proposed a new analytical fitting function able to model the spectra and iii) tested an innovative experimental protocol, Polarization Gated Brillouin Spectroscopy (PG-BS), able to subtract multiple scattering without recurring to a sophisticated data analysis.

We believe that, in the next future, the combined use of the analytic and the experimental methods here proposed can extend the application of Brillouin analysis to new classes of heterogeneous materials. In particular, PG-BS, with minimal computational effort, can be universally applicable to scattering media regardless of their constitution, scattering strength and composition. Moreover, the ability to detect objects of different stiffness hidden inside turbid media, constituting the most of the biological materials, paves the way for the translation of BM to new applications for pathology diagnosis.

*Biomechanics of Intracellular Stress Granules by ALS Protein FUS*, Communications Biology **1**, 139 (2018).

[13] S. Mattana, M. Mattarelli, L. Urbanelli, K. Sagini, C. Emiliani, M. D. Serra, D. Fioretto, and S. Caponi, *Non-Contact Mechanical and Chemical Analysis of Single Living Cells by Microspectroscopic Techniques*, Light: Science & Applications **7**, 17139 (2018).

[14] I. V. Kabakova, Y. Xiang, C. Paterson, and P. Török, *Fiber-Integrated Brillouin Microspectroscopy: Towards Brillouin Endoscopy*, Journal of Innovative Optical Health Sciences **10**, 1742002 (2017).

[15] R. R. Alfano, W. B. Wang, L. Wang, and S. K. Gayen, *Light Propagation in Highly Scattering Turbid Media: Concepts, Techniques, and Biomedical Applications*, in *Photonics*, Vol. IV (Wiley, 2015), pp. 367–412.

[16] A. Yaroshevsky, Z. Glasser, E. Granot, and S. Sternklar, *Transition from the Ballistic to the Diffusive Regime in a Turbid Medium*, Optics Letters **36**, 1395 (2011).

[17] F. Scarponi, S. Mattana, S. Corezzi, S. Caponi, L. Comez, P. Sassi, A. Morresi, M. Paolantoni, L. Urbanelli, C. Emiliani, L. Roscini, L. Corte, G. Cardinali, F. Palombo, J. R. Sandercock, and D. Fioretto, *High-Performance Versatile Setup for Simultaneous Brillouin-Raman Microspectroscopy*, Physical Review X **7**, 031015 (2017).

[18] G.-O. Regnima, T. Koffi, O. Bagui, A. Kouacou, E. Kristensson, J. Zoueu, and E. Berrocal, *Quantitative Measurements of Turbid Liquids via Structured Laser Illumination Planar Imaging Where Absorption Spectrophotometry Fails*, Applied Optics **56**, 3929 (2017).

[19] E. Berrocal, D. L. Sedarsky, M. E. Paciaroni, I. V. Meglinski, and M. A. Linne, *Laser Light Scattering in Turbid Media Part I: Experimental and Simulated Results for the Spatial Intensity Distribution*, Optics Express **15**, 10649 (2007).

[20] K. M. Yoo and R. R. Alfano, *Time-Resolved Coherent and Incoherent Components of Forward Light Scattering in Random Media*, Optics Letters **15**, 320 (1990).

[21] R. Mercatelli, S. Mattana, L. Capozzoli, F. Ratto, F. Rossi, R. Pini, D. Fioretto, F. S. Pavone, S. Caponi, and R. Cicchi, *Morpho-Mechanics of Human Collagen Superstructures Revealed by All-Optical Correlative Micro-Spectroscopies*, Communications Biology **2**, 117 (2019).

[22] D. Fioretto, S. Caponi, and F. Palombo, *Brillouin-Raman Mapping of Natural Fibers with Spectral Moment Analysis*, Biomedical Optics Express **10**, 1469 (2019).

[23] M. Pochylski and J. Gapiński, *Simple Way to Analyze Brillouin Spectra from Turbid Liquids*, Optics Letters **40**, 1456 (2015).

[24] S. Corezzi, L. Comez, and M. Zanatta, *A Simple Analysis of Brillouin Spectra from Opaque Liquids and Its Application to Aqueous Suspensions of Poly-N-Isopropylacrylamide Microgel Particles ☆*, Journal of Molecular Liquids **266**, 460 (2018).

[25] L. WANG, P. P. HO, C. LIU, G. ZHANG, and R. R. ALFANO, *Ballistic 2-D Imaging Through Scattering Walls Using an Ultrafast Optical Kerr Gate*, Science **253**, 769 (1991).

[26] S. G. Demos and R. R. Alfano, *Optical Polarization Imaging*, Applied Optics **36**, 150 (1997).

[27] H. Z. Cummins, G. Li, W. Du, R. M. Pick, and C. Dreyfus, *Origin of Depolarized Light Scattering in Supercooled Liquids: Orientational Fluctuation versus Induced Scattering Mechanisms*, Physical Review E **53**, 896 (1996).

Supplementary materials:

Brillouin imaging in turbid samples: the removal of multiple scattering contribution

Maurizio Mattarelli[1], Giulio Capponi[1], Alessandra Anna Passeri[1], Daniele Fioretto[1], Silvia Caponi [2]*

[1] Dipartimento di Fisica e Geologia, Università di Perugia, Via A. Pascoli, I-06100 Perugia, Italy

[2] Istituto Officina dei Materiali, Italian National Research Council (IOM-CNR), Unit of Perugia, c/o Department of Physics and Geology, University of Perugia, Via A. Pascoli, I-06123 Perugia, Italy

In a scattering event, the scattering angle θ is defined as the angle between $\mathbf{k}_i$ and $\mathbf{k}_f$, the wave-vectors of the incoming and the scattered photon respectively, as illustrated in Fig. 1s.

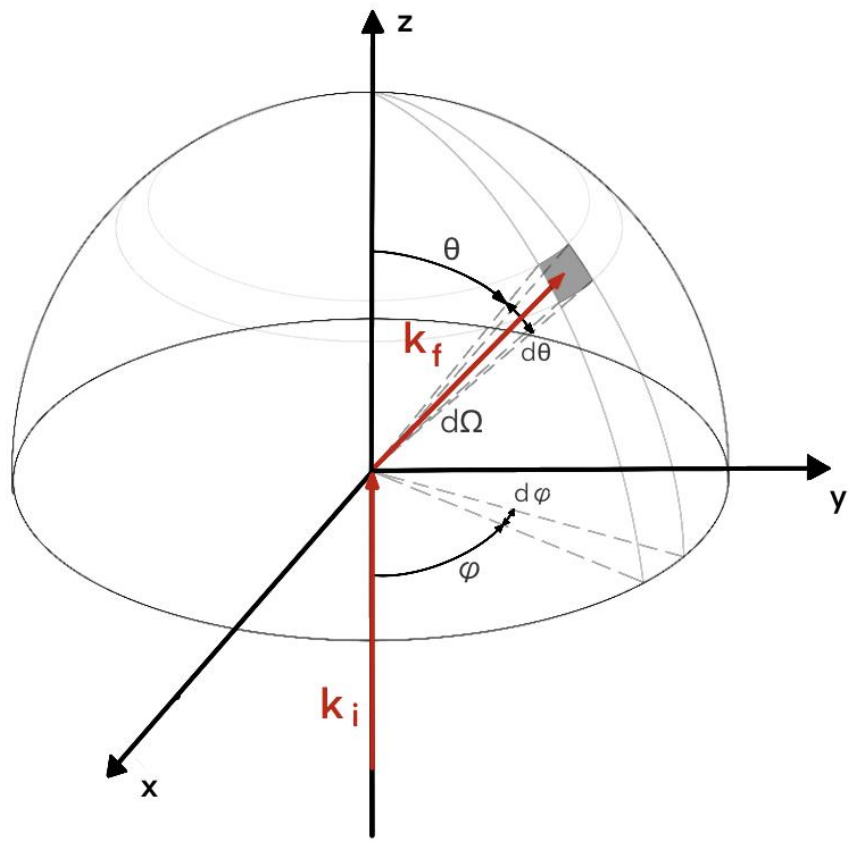

Figure 1S: The incoming, $\mathbf{k}_i$, and the scattered, $\mathbf{k}_f$, wave-vectors of photons participating to a Brillouin scattering event are represented on a sphere of radius $\|\mathbf{k}_f\|=\|\mathbf{k}_i\|$. The scattering angle θ is indicated.

Considering an isotropic distribution of the scattered radiation, the probability to find $\mathbf{k_f}$ at a certain orientation with respect to $\mathbf{k_i}$, fixed along the z axis, is proportional to the solid angle $d\Omega$

$$d\Omega = d\cos(\theta)\, d\varphi$$

In Brillouin Spectroscopy, it is more useful to express this quantity by the exchanged wavevector $q = \frac{4 n \pi \sin\left(\frac{\theta}{2}\right)}{\lambda}$, where $n$ is the refractive index of the sample and $\lambda$ is the light wavelength.

Then the expression for the solid angle takes the form:

$$d\Omega = d\cos(\theta)\, d\varphi = \sin(\theta)\, d\theta d\varphi = 4\sin\left(\frac{\theta}{2}\right)\cos\left(\frac{\theta}{2}\right) d\left(\frac{\theta}{2}\right) d\varphi = 4\sin\left(\frac{\theta}{2}\right) d\left(\sin\left(\frac{\theta}{2}\right)\right) d\varphi \propto q\, dq\, d\varphi \quad (1s)$$

This expression can be exploited to evaluate the intensity of Brillouin scattering acquired in turbid media, that is due to the scattering events occurring to ballistic and diffusive photons, $I_{BS}$ and $I_{MS}$ respectively.

The total scattered intensity can be written a sum of the two contributions, which in presence of a distribution of scattering wave-vectors take the form:

$$I(\omega) = I_{BS}(\omega) + I_{MS}(\omega) = A_{BS}\int_{Q_{obj}} \frac{(vq)^2 Dq^2}{(\omega^2-(vq)^2)+\omega^2(Dq^2)^2} R(\boldsymbol{q})\, dq +$$
$$A_{MS}\int_0^{q_{180}} \frac{(vq)^2 Dq^2}{(\omega^2-(vq)^2)+\omega^2(Dq^2)^2} R_{MS}(q)\, dq \quad (2s)$$

The first integral takes into account the q spread related to the numerical aperture of the microscope objective, $Q_{obj}$ The second integral takes into account the complete loss of q definition due to multiple scattering.

In order to evaluate the integrals, it has been considered that, for an isotropic distribution of the scattered radiation, the distribution probability is proportional to the solid angle, i.e $R_{MS}(q)dq \propto d\Omega \propto qdq$.

Moreover, in backscattering configuration also $R(\boldsymbol{q}) \propto q$, if we can assume that the q spread is only in the collection of the scattered radiation and the objective transmission does not depend on the angle. In this case, as illustrated in Figure 2S, the integration limits of the first integral are $q_{min}$ and $q_{180}$ respectively, where $q_{180}$ is the exchange wavevector in back scattering configuration and $q_{min} = q_{180}$ $\sin\left(\frac{\theta_{min}}{2}\right)$, where $\theta_{min} = \pi - \alpha$, $\alpha$ being the maximum acceptance angle of the microscope objective. Note that if there is a refractive index mismatch between the sample and the medium, then the minimum angle relevant for Brillouin scattering has to be corrected for refraction according to the Snell's law.

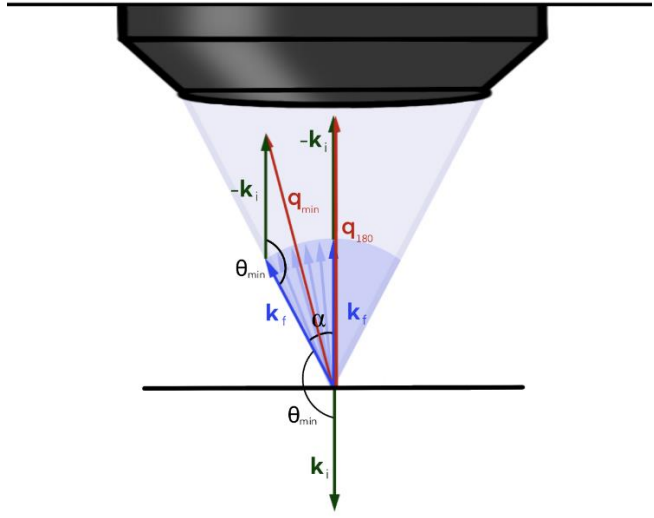

Apart from the integration limits we can see that both $I_{MS}$ and $I_{BS}$ can be described by the following expression

$$F(\omega) = \int \frac{(vq)^2 Dq^2}{(\omega^2 - (vq)^2)^2 + \omega^2 (Dq^2)^2} q \, dq \quad (3s)$$

Eq (3s) has a closed form expression, which, even if cumbersome, is well suited to be used in fitting procedures:

$$F(\omega) = \frac{\left(v^2\left(\omega(v^4 - D^2\omega^2)\tan^{-1}\left(\frac{q^2(D^2\omega^2 + v^4) - v^2\omega^2}{D\omega^3}\right) + D\left(q^2(D^2\omega^2 + v^4) + v^2\omega^2 \log\left(\frac{q^4(D^2\omega^2 + v^4) - 2q^2v^2\omega^2 + \omega^4}{v^4 q_{BS}^4}\right)\right)\right)\right)}{2\pi(D^2\omega^2 + v^4)^2}$$

The number of fitting parameters, apart from the scale factor and an eventual background is therefore 3: the two parameters, $v$ and $D$, describing the frequency dependent complex elastic modulus $M^*(\omega) = \rho v^2 + i \rho D \omega$, and the ratio $A_{BS}/A_{MS}$.